\documentclass[aps,prb,twocolumn,showpacs,amssymb,amsmath]{revtex4-1}
\usepackage{prettyref}
\usepackage{graphicx}
\usepackage{verbatim}   
\usepackage{color}      
\usepackage{subfigure}  
\usepackage{hyperref}

\usepackage{bm}
\usepackage{epsfig,amsopn}
\usepackage{amsmath,amssymb}
\usepackage{natbib}

\begin{document}

\title{Cumulants of heat transfer across nonlinear quantum systems}

\author{Huanan Li$^{1}$}
\author{Bijay Kumar Agarwalla$^{1}$}
\author{Baowen Li$^{1,2,3}$}
\author{Jian-Sheng Wang$^{1}$}

\affiliation{$^{1}$ Department of Physics and Center for Computational Science and Engineering,
National University of Singapore, Singapore 117542, Republic of Singapore\\
$^{2}$ NUS Graduate School for Integrative Sciences and Engineering, Singapore 117456, Republic of Singapore\\
$^{3}$ NUS-Tongji Center for Phononics and Thermal Energy Science, School of Physical Science and Engineering, Tongji University, 200092 Shanghai, China}

\date{25 Aug 2012}
\begin{abstract}
We consider thermal conduction across a general nonlinear phononic
junction. Based on two-time observation protocol and the field theoretical/algebraic
method, the cumulants of the heat transferred in both transient and steady-state
regimes are studied on an equal footing, and practical formulae for the calculation of the
cumulant generating function of heat transfer are obtained. As an application, the developed general formalism is used to study anharmonic effects on
fluctuation of steady-state heat transfer across a single-site
junction with a quartic nonlinear on-site pinning potential.
An explicit nonlinear modification to cumulant generating function
exact up to the first order is given, in which Gallavotti-Cohen fluctuation
symmetry is verified. Numerically a self-consistent
procedure is introduced, which works well for strong nonlinearity.
\end{abstract}


\maketitle

\section{INTRODUCTION}

The physics of nonequilibrium many-body systems is one of the most
rapidly expanding areas which attracts much attentions recently. With
the development of the modern nanoscale technology, a full understanding
of the general features of thermal conduction is needed. And it is
well-known that the noise \cite{Blanter2000} generated by nanodevices
contains valuable information on microscopic transport processes not
available from current, which is simply related to the lowest order
moment of the full distribution function for heat transfer. To this
end, the full counting statistics (FCS) regarding the distribution
of transferred quantity, such as heat in phononic case, has to be
determined.

The extensive study of the FCS started from the field of electronic
transport pioneered by Levitov and Lesovik, \cite{Levitov1993} while
much less attention is given to heat transfer. Saito and Dhar are
the first ones to borrow this concept to thermal transport. \cite{Saito2007}
However, both of them, including many subsequent works, are mainly
restricted to noninteracting systems,~\cite{Schonhammer2007,Wang2011}
although some works have already been devoted to the analysis of fluctuation
considering the effect of nonlinearity in the classical limit through
Langevin simulations, or approximately in a restricted electronic
transport case, such as the FCS in molecular junctions with electron-phonon
interaction. \cite{Avriller2009}

In recent years, phononics, i.e., the counterpart technology of electronics,
has great interest to both theorists and experimentalists, and presents
an unforseen wealth of applications.~\cite{Wu2009,Li2012} And the nonlinearity, such as
the phonon-phonon interaction, has been found of special importance
in the construction of phononic devices.~\cite{Li2012} In addition,
recently it has been noted that the nonlinearity of interaction is
crucial to the manifestation of geometric heat flux.~\cite{Ren2012}
Therefore, it is desirable to establish a systematic and practical
formalism to properly deal with cumulants of heat transfer in the
presence of nonlinearity.

In this work, we shall study the FCS for heat transfer flowing across
a quantum junction in the presence of general phonon-phonon interactions.
Then, based on two-time observation protocol,~\cite{Esposito2009}
we construct a concise and rigorous cumulant generating function (CGF)
expression for the heat transfer in this general situation, in which
both the transient and steady-state case are dealt with on equal footing.
Furthermore, as an illustration of this general formalism, a single-site junction with a quartic nonlinear on-site pinning potential is
introduced and the corresponding CGF exact up to first order of the
nonlinear interaction strength is given. Finally, a self-consistent
scheme is employed to numerically illustrate the nonlinear effects
on first three cumulants of heat transfer. In short, the paper
is organized as follows. We introduce the model and elaborate the general formalism
in Sec.~\ref{sec:MODEL-AND-FORMALISM}, which is taken as the main part of the work. And in Sec.~\ref{sec:APPLICATION},
an application is introduced to verify the general formalism. Finally we summarize in Sec.~\ref{sec:SUMMARY}.

\section{MODEL AND THE GENERAL FORMALISM\label{sec:MODEL-AND-FORMALISM}}
We consider the lead-junction-lead model initially prepared in a product
state $\rho_{ini}=\Pi_{\alpha=L,C,R}\frac{e^{-\beta_{\alpha}H_{\alpha}}}{\mathrm{Tr}\left(e^{-\beta_{\alpha}H_{\alpha}}\right)}$.
It can be imagined that left lead $\left(L\right)$, center junction
$\left(C\right)$, and right lead $\left(R\right)$ in this model
were in contact with three different heat baths at the inverse temperatures
$\beta_{L}=\left(k_{B}T_{L}\right)^{-1}$, $\beta_{C}=\left(k_{B}T_{C}\right)^{-1}$
and $\beta_{R}=\left(k_{B}T_{R}\right)^{-1}$, respectively, for time
$t<t_{0}$. At time $t=t_{0}$, all the heat baths are removed, and
couplings of the center junction with the leads $H_{LC}=u_{L}^{T}V^{LC}u_{C}$
and $H_{CR}=u_{C}^{T}V^{CR}u_{R}$ and the interested nonlinear term
$H_{n}$ appearing only in the center junction are switched on
abruptly. Now the total Hamiltonian is given by
\begin{align}
H_{tot}= & \:H_{L}+H_{C}+H_{R}+H_{LC}+H_{CR}+H_{n},\label{eq:total H}
\end{align}
where $H_{\alpha}=\frac{1}{2}p_{\alpha}^{T}p_{\alpha}+$$\frac{1}{2}u_{\alpha}^{T}K^{\alpha}u_{\alpha},\;\alpha=L,C,R$,
represents coupled harmonic oscillators, $u_{\alpha}=\sqrt{m_{\alpha}}x_{\alpha}$
and $p_{\alpha}$ are column vectors of transformed coordinates and
corresponding conjugate momenta in region $\alpha$. $K^\alpha$ is force constant matrix; the superscript
$T$ stands for matrix transpose.

We can construct a consistent framework~\cite{Griffiths2002} consisting of two-time
quantum histories to study the heat transfer across arbitrary nonlinear systems in a given time duration.~\cite{Li_CGF}
A realization of the quantum history $P_{t_{0}}^{a}\odot P_{t_{M}}^{b}$ means that the result of the measurement at time $t_{0}$ of energy
of the left lead associated with the operator $H_{L}$ is
the eigenvalue $a$ of $H_{L}$, then the measurement at
time $t_{M}$ yields the eigenvalue $b$ of $H_{L}$.
Using this consistent quantum framework, we can define the generating function (GF) for heat transfer in the time duration  $t_{M}-t_{0}$ to be
\begin{align}
\mathcal{Z}\left(\xi\right)\equiv & \sum_{a,b}e^{i\left(a-b\right)\xi}\Pr\left(P_{t_{0}}^{a}\odot P_{t_{M}}^{b}\right)\nonumber \\
= & \left\langle U_{\xi/2}\left(t_{0},t_{M}\right)U_{-\xi/2}\left(t_{M},t_{0}\right)\right\rangle ,
\end{align}
where $\Pr\left(P_{t_{0}}^{a}\odot P_{t_{M}}^{b}\right)$ stands
for the joint probability for the quantum history $P_{t_{0}}^{a}\odot P_{t_{M}}^{b}$;  $U_{-\xi/2}\left(t_{M},t_{0}\right)$
means evolution operator associated with the counting-field dependent total Hamiltonian
$H_{tot}^{-\xi/2}=e^{i\left(-\xi/2\right)H_{L}}H_{tot}e^{-i\left(-\xi/2\right)H_{L}}$,
similarly for $U_{\xi/2}\left(t_{0},t_{M}\right)$ and $\left\langle \ldots\right\rangle $ denotes the ensemble average over the initial state $\rho_{ini}$.

\begin{figure}[h!]
\includegraphics[width=3.2in,clip]{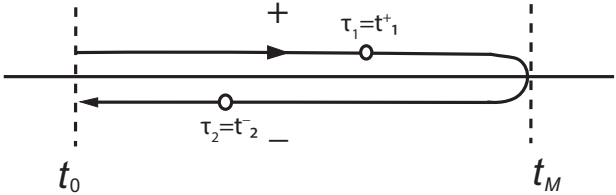}
\caption{An illustration of the contour $C$. The upper branch is called $+$ and lower one $-$ so that
a particular time point $\tau_1$ on the upper branch is denoted by $t_1^+$
while $\tau_2$ on the lower one by $t_2^-$. The time order follows the direction of the arrows.
\label{Contour}}
\end{figure}
The first step for the study of the GF is to relate it to the Green's function, by which the closed equation satisfied can be found out.
To this end, we generalize the GF to be
\begin{eqnarray}
& &\mathcal{Z}\left(\lambda_{2}-\lambda_{1}\right) \nonumber \\
  &\equiv& \left\langle U_{\lambda_{2}}\left(t_{0},t_{M}\right)U_{\lambda_{1}}\left(t_{M},t_{0}\right)\right\rangle \\
   &=&\left\langle e^{i\left(\lambda_2-\lambda_1\right)H_L}U\left(t_{0},t_{M}\right)e^{-i\left(\lambda_2-\lambda_1\right)H_L}U\left(t_{M},t_{0}\right)\right\rangle\\
   &=&\left\langle T_{\tau}e^{-\frac{i}{\hbar}\int_{C}d\tau\mathcal{T}_{\lambda}\left(\tau\right)}\right\rangle,
\end{eqnarray}
where, $T_{\tau}$ is a $\tau$-ordering super-operator arranging operators
with earlier $\tau$ on the contour $C$ (from $t_{0}$ to $t_{M}$
and back to $t_{0}$) to the right, see Fig.~\ref{Contour}. In the second equality we have used the cyclic property of the trace and the commutator relation $[H_L,\,\rho_{ini}]=0$;
in the third equality we
go to the interaction picture with respect to the free Hamiltonian $h=H_{L}+H_{C}+H_{R}$ so that
$\mathcal{T}_{\lambda}\left(\tau\right)=\hat{u}_{L}^{x,T}\left(\tau\right)V^{LC}\hat{u}_{C}\left(\tau\right)+\hat{u}_{C}^{T}\left(\tau\right)V^{CR}\hat{u}_{R}\left(\tau\right)+\hat{\mathcal{H}}_{n}\left(\tau\right)$
with caret put above operators to denote the interaction-picture $\tau$ dependence such as $\hat{u}_{C}\left(\tau\right)=e^{\frac{i}{\hbar}h\tau}u_{C}e^{-\frac{i}{\hbar}h\tau}$),
where $\hat{u}_{L}^{x}\left(\tau\right)=\hat{u}_{L}\left(\hbar x_{\tau}+\tau\right)$
with $x_{\tau}=\lambda_{1}$ ($\lambda_{2}$) with $\tau=t^{+}$ ($t^{-}$)
on the upper (lower) branch of the contour $C$.

Furthermore, we define the adiabatic potential
$\mathcal{U}\left(t,\lambda_{2},\lambda_{1}\right)$
according to~\cite{Gogolin2006}
\begin{equation}
\mathcal{Z}\left(\lambda_{2}-\lambda_{1}\right)=e^{-\frac{i}{\hbar}\int_{t_{0}}^{t_{M}}dt\,\mathcal{U}\left(t,\lambda_{2},\lambda_{1}\right)}.
\end{equation}
Thus we could apply the nonequilibrium version of the Feynman-Hellmann
theorem \cite{Feynman1939} to get
\begin{align}
\frac{\partial}{\partial\lambda_{1}}\mathcal{U}\left(t,\lambda_{2},\lambda_{1}\right)= & \frac{1}{\mathcal{Z}\left(\lambda_{2}-\lambda_{1}\right)}\left\langle T_{\tau}\frac{\partial\mathcal{T}_{\lambda}\left(t^{+}\right)}{\partial\lambda_{1}}e^{-\frac{i}{\hbar}\int_{C}d\tau\mathcal{T}_{\lambda}\left(\tau\right)}\right\rangle \\
\equiv & \left\langle T_{\tau}\frac{\partial\mathcal{T}_{\lambda}\left(t^{+}\right)}{\partial\lambda_{1}}\right\rangle _{\lambda}.\label{eq: Feynman-Hellmann theorem}
\end{align}

Since
\begin{equation}
\frac{\partial\mathcal{T}_{\lambda}\left(t^{+}\right)}{\partial\lambda_{1}}=\hbar\frac{\partial\hat{u}_{L}^{x,T}\left(t^{+}\right)}{\partial t^{+}}V^{LC}\hat{u}_{C}\left(t^{+}\right)
\end{equation}
and introducing contour-ordered Green's functions $\tilde{G}_{LC}$ and $\tilde{G}_{CL}$ defined as
\begin{eqnarray}
\tilde{G}_{LC}\left(\tau_{1},\,\tau_{2}\right) & = & -\frac{i}{\hbar}\left\langle T_{\tau}\hat{u}_{L}^{x}\left(\tau_{1}\right)\hat{u}_{C}^{T}\left(\tau_{2}\right)\right\rangle _{\lambda}\\
\tilde{G}_{CL}\left(\tau_{1},\,\tau_{2}\right) & = & -\frac{i}{\hbar}\left\langle T_{\tau}\hat{u}_{C}\left(\tau_{1}\right)\hat{u}_{L}^{x,T}\left(\tau_{2}\right)\right\rangle _{\lambda},
\end{eqnarray}
we can get
\begin{eqnarray}
&&\frac{\partial\ln\mathcal{Z}\left(\lambda_{2}-\lambda_{1}\right)}{\partial\lambda_{1}}\nonumber\\
 & = & \int_{t_{0}}^{t_{M}}dt\hbar\frac{\partial}{\partial t'}\mathrm{Tr}\left[\tilde{G}_{LC}^{t}\left(t',\, t\right)V^{CL}\right]\bigg|_{t'=t}\nonumber\\
 & = & \int_{t_{0}}^{t_{M}}dt\hbar\frac{\partial}{\partial t'}\mathrm{Tr}\left[\tilde{G}_{CL}^{t}\left(t,\, t'\right)V^{LC}\right]\bigg|_{t'=t}.\label{partial_lnZ_first}
\end{eqnarray}
Notice that the tilde on the Green's functions emphasizes
the fact that they are counting field $\xi$-dependent, and
real-time Green's functions can be obtained by specifying the variation range of the
time arguments in contour-ordered Green's functions such as
\begin{eqnarray*}
  \tilde{G}_{LC}\left(\tau_{1},\tau_{2}\right) &\rightarrow& \begin{bmatrix}\tilde{G}_{LC}\left(t_{1}^{+},t_{2}^{+}\right) & \tilde{G}_{LC}\left(t_{1}^{+},t_{2}^{-}\right)\\
\tilde{G}_{LC}\left(t_{1}^{-},t_{2}^{+}\right) & \tilde{G}_{LC}\left(t_{1}^{-},t_{2}^{-}\right)
\end{bmatrix}\\
   &=& \begin{bmatrix}\tilde{G}_{LC}^{t}\left(t_{1},t_{2}\right) & \tilde{G}_{LC}^{<}\left(t_{1},t_{2}\right)\\
\tilde{G}_{LC}^{>}\left(t_{1},t_{2}\right) & \tilde{G}_{LC}^{\bar{t}}\left(t_{1},t_{2}\right)
\end{bmatrix}.
\end{eqnarray*}
According to the basic analysis of Feynman diagrams,~\cite{Rammer2007} the contour-ordered Green's functions $\tilde{G}_{LC}$ and $\tilde{G}_{CL}$
are given as
\begin{eqnarray}
\tilde{G}_{LC}\left(\tau_{1},\tau_{2}\right) & = & \int_{C}\tilde{g}^{L}\left(\tau_{1},\tau\right)V^{LC}\tilde{G}_{CC}\left(\tau,\tau_{2}\right)d\tau, \label{G_LC}\\
\tilde{G}_{CL}\left(\tau_{1},\tau_{2}\right) & = & \int_{C}\tilde{G}_{CC}\left(\tau_{1},\tau\right)V^{CL}\tilde{g}^{L}\left(\tau,\tau_{2}\right)d\tau, \label{G_CL}
\end{eqnarray}
with the shifted bare Green's function for the left lead being
\begin{equation}
\tilde{g}^{L}\left(\tau_{1},\tau_{2}\right)=-\frac{i}{\hbar}\left\langle T_{\tau}\hat{u}_{L}^{x}\left(\tau_{1}\right)\hat{u}_{L}^{x,T}\left(\tau_{2}\right)\right\rangle,
\end{equation}
where
\begin{equation}
\tilde{G}_{CC}\left(\tau_{1},\tau_{2}\right)=-\frac{i}{\hbar}\left\langle T_{\tau}\hat{u}_{C}\left(\tau_{1}\right)\hat{u}_{C}^{T}\left(\tau_{2}\right)\right\rangle _{\lambda} \label{GCC_def}
\end{equation}
is the central quantity for the study of the GF of the heat transfer, which we will discuss later.
Before that we use the treatment of symmetrization to simplify $\frac{\partial\ln\mathcal{Z}}{\partial\lambda_{1}}$ in Eq.~\eqref{partial_lnZ_first} further
according to the time-order version of Eq.~\eqref{G_LC} and Eq.~\eqref{G_CL}, \textit{i.e.},
\begin{eqnarray}
\tilde{G}_{LC}^{t}\left(t',t\right) & = & \int_{t_{0}}^{t_{M}}\tilde{g}_{L}^{t}\left(t',t_{1}\right)V^{LC}\tilde{G}_{CC}^{t}\left(t_{1},t\right)dt_{1}\nonumber \\
&-&\int_{t_{0}}^{t_{M}}\tilde{g}_{L}^{<}\left(t',t_{1}\right)V^{LC}\tilde{G}_{CC}^{>}\left(t_{1},t\right)dt_{1} \\
\tilde{G}_{CL}^{t}\left(t,t'\right) & = & \int_{t_{0}}^{t_{M}}\tilde{G}_{CC}^{t}\left(t,t_{1}\right)V^{CL}\tilde{g}_{L}^{t}\left(t_{1},t'\right)dt_{1}\nonumber\\
&-&\int_{t_{0}}^{t_{M}}\tilde{G}_{CC}^{<}\left(t,t_{1}\right)V^{CL}\tilde{g}_{L}^{>}\left(t_{1},t'\right)dt_{1},
 \end{eqnarray}
%
which explicitly means that
\begin{align}
 &\frac{\partial\ln\mathcal{Z}\left(\lambda_{2}-\lambda_{1}\right)}{\partial\lambda_{1}}\nonumber\\
= & \frac{\hbar}{2}\int_{t_{0}}^{t_{M}}dt\frac{\partial}{\partial t'}\mathrm{Tr}\Big[\tilde{G}_{CL}^{t}\left(t,t'\right)V^{LC}+\tilde{G}_{LC}^{t}\left(t',t\right)V^{CL}\Big]\bigg|{}_{t'=t}\\
= & -\frac{\hbar}{2}\int_{t_{0}}^{t_{M}}dtdt'\mathrm{Tr}\Big[\tilde{G}_{CC}^{>}\left(t,t'\right)\frac{\partial\tilde{\Sigma}_{L}^{<}\left(t',t\right)}{\partial t'}\nonumber \\
 & +\tilde{G}_{CC}^{<}\left(t,t'\right)\frac{\partial\tilde{\Sigma}_{L}^{>}\left(t',t\right)}{\partial t}\Big]\label{eq:Meir-Wingeen}
\end{align}
with the self-energy defined to be $\tilde{\Sigma}_{L}\left(\tau_{1},\tau_{2}\right)=V^{CL}\tilde{g}_{L}\left(\tau_{1},\tau_{2}\right)V^{LC}$.
Eq.~\eqref{eq:Meir-Wingeen} is a generalized Meir-Wingreen formula.
In obtaining the second equality we have used the relation $\frac{\partial\tilde{\Sigma}_{L}^{t}\left(t',t_{1}\right)}{\partial t'}=-\frac{\partial\tilde{\Sigma}_{L}^{t}\left(t',t_{1}\right)}{\partial t_{1}}$ since $\tilde{\Sigma}_{L}^{t}\left(t',t_{1}\right)=\tilde{\Sigma}_{L}^{t}\left(t'-t\right)$.
Essentially we employ the procedure of symmetrization to get rid of
the time-ordered version of $\tilde{G}_{CC}\left(\tau_{1},\tau_{2}\right)$, \textit{i.e.}, $\tilde{G}_{CC}^{t}\left(t_{1},t_{2}\right)$.

Setting $\lambda_{1}=-\xi/2$ and $\lambda_{2}=\xi/2$, and noticing that
$\frac{\partial\tilde{\Sigma}_{L}^{<,>}\left(t',t_{1}\right)}{\partial t'}  =  -\frac{1}{\hbar}\frac{\partial\tilde{\Sigma}_{L}^{<,>}\left(t',t_{1}\right)}{\partial\xi}$
and $\frac{\partial\tilde{\Sigma}_{L}^{t,\bar{t}}\left(t',t_{1}\right)}{\partial\xi}  =  0$,
we can obtain a compact expression for $\frac{\partial\ln\mathcal{Z}}{\partial\left(i\xi\right)}$
from the generalized Meir-Wingreen formula Eq.~\eqref{eq:Meir-Wingeen}:
\begin{eqnarray}
\frac{\partial\ln\mathcal{Z}}{\partial\left(i\xi\right)} & = & \frac{1}{2}\int_{t_{0}}^{t_{M}}dt\int_{t_{0}}^{t_{M}}dt'\mathrm{Tr}\bigg \{ \begin{pmatrix}\tilde{G}_{CC}^{t}\left(t,t'\right) & \tilde{G}_{CC}^{<}\left(t,t'\right)\\
-\tilde{G}_{CC}^{>}\left(t,t'\right) & -\tilde{G}_{CC}^{\bar{t}}\left(t,t'\right)
\end{pmatrix}\nonumber\\
&&\begin{pmatrix}0 & \frac{\partial\tilde{\Sigma}_{L}^{<}\left(t',t\right)}{\partial\left(i\xi\right)}\\
-\frac{\partial\tilde{\Sigma}_{L}^{>}\left(t',t\right)}{\partial\left(i\xi\right)} & 0
\end{pmatrix}\bigg \} \nonumber\\
 & = & \frac{1}{2}\int_{C}d\tau\int_{C}d\tau'\mathrm{Tr}\left[\tilde{G}_{CC}\left(\tau,\tau'\right)\frac{\partial\tilde{\Sigma}_{L}\left(\tau',\tau\right)}{\partial\left(i\xi\right)}\right].
\label{generalized Meir_wingreen formula}
\end{eqnarray}
It is worth mentioning that, Eq.~\eqref{generalized Meir_wingreen formula} could
be also obtained based on the field theoretical/diagrammatic method.~\cite{Kleinert2000,Pelster2004} If needed,
the proper normalization for the CGF, $i.e.$, $\ln\mathcal{Z}\left(\xi\right)$,
can be determined by the constraint $\ln\mathcal{Z}\left(0\right)=0$.

\section{\label{Interaction picture on the contour}Picture on the contour}
The nonlinear effects on the GF are completely included in the $\tilde{G}_{CC}$, for
which we try to obtain the closed Dyson equations now. To this end, we need to introduce \textit{the picture on the contour}.
A key concept appearing in \textit{the picture on the contour} is an evoulution operator $U^{S}\left(\tau_{2},\,\tau_{1}\right)$ defined on the contour $C$.
 Assuming that $\tau_{2}\succ\tau_{1}$, namely $\tau_2$ succeeds $\tau_1 $ on the contour, we will encounter three different situations depending on the relative position of the arguments $\tau_{2}$ and $\tau_{1}$:
\begin{eqnarray}
U^{S}\left(\tau_{2},\,\tau_{1}\right)\quad\quad\quad\quad\quad\quad\quad\quad\quad\quad\quad\quad\quad\quad\quad\quad\quad\quad&& \nonumber\\
=\begin{cases}
U_{S}^{+}\left(t_{2},\, t_{1}\right), & \left(\tau_{2}=t_{2}^{+}\right)>\left(\tau_{1}=t_{1}^{+}\right)\\
U_{S}^{-}\left(t_{2},\, t_{M}\right)U_{S}^{+}\left(t_{M},\, t_{1}\right), & \tau_{2}=t_{2}^{-},\,\tau_{1}=t_{1}^{+}\\
U_{S}^{-}\left(t_{2},\, t_{1}\right). & \left(\tau_{2}=t_{2}^{-}\right)<\left(\tau_{1}=t_{1}^{-}\right)
\end{cases}.\nonumber\\
&&\label{evolution operator on the contour}
\end{eqnarray}
Where in the second situation we need not specify the relative magnitude of $t_2^-$ and $t_1^+$, since the time on the
lower branch always succeeds the time on the upper branch along the contour. Also we should notice that the superscript $+$
or $-$ for the evolution operator simply tell us that the ordinary Schr\"{o}dinger evolution operator is for the upper branch
or the lower branch, respectively.  Compactly, the evolution operator
defined on the contour $U^{S}\left(\tau_{2},\,\tau_{1}\right)$ when $\tau_{2}\succ\tau_{1}$ could be written as
\begin{equation}
    U^{S}\left(\tau_{2},\,\tau_{1}\right)=T_{\tau}\exp\left(-\frac{i}{\hbar}\int_{C\left[\tau_{2},\,\tau_{1}\right]}H_{tot}\left(\tau\right)d\tau\right),
\end{equation}
where $C[\tau_{1},\tau_{2}]$ denotes the path along the contour $C$ from $\tau_{2}$
to $\tau_{1}$. In order to keep group properties of the evolution operator, the evolution operator $U^{S}\left(\tau_{1},\,\tau_{2}\right)$ when $\tau_{2}\succ\tau_{1}$
is defined to be
\begin{equation}
 U^{S}\left(\tau_{1},\,\tau_{2}\right)=U^{S}\left(\tau_{2},\,\tau_{1}\right)^{-1}=U^{S}\left(\tau_{2},\,\tau_{1}\right)^{\dagger}.
\end{equation}

After this general discussion, we will use the same notation $U^{S}\left(\tau_{2},\,\tau_{1}\right)$ to denote the Schr\"{o}dinger-picture evolution operator on the contour without causing confusion, which is determined by the effective total Hamiltonian $H_{tot}^{x}\left(\tau\right)\equiv e^{ix_\tau H_{L}}H_{tot}e^{-ix_\tau H_{L}}$. Then the Heisenberg-picture operator on the contour such as
$u_{C}^{H}\left(\tau_{1}\right)$ is defined as
\begin{equation}
u_{C}^{H}\left(\tau_{1}\right)=U^{S}\left(t_{0}^{+},\,\tau_{1}\right)u_{C}U^{S}\left(\tau_{1},\, t_{0}^{+}\right).
\end{equation}
By virtue of the Heisenberg-picture on the contour,
we can rewrite the $\tilde{G}_{CC}$ from Eq.~\eqref{GCC_def} as
\begin{eqnarray}
   && \tilde{G}_{CC}\left(\tau_{1},\tau_{2}\right)= \nonumber\\
   &&-\frac{i}{\hbar}\left\langle U^{S}\left(t_{0}^{-},t_{M}^{-}\right)U^{S}\left(t_{M}^{+},t_{0}^{+}\right)T_{\tau}u_{C}^{H}\left(\tau_{1}\right)u_{C}^{H,T}\left(\tau_{2}\right)\right\rangle \frac{1}{\mathcal{Z}}.\nonumber\\
   &&\label{Heisenberg_picture_G_CC}
\end{eqnarray}

Now we define the interaction picture on the contour.
The modified total Hamiltonian can be split into two parts, \textit{i.e.},
\begin{equation}
H_{tot}^{x}\left(\tau\right)=H_{0}^{x}\left(\tau\right)+H_{n}.
\end{equation}
The interaction-picture evolution operator is defined as
\begin{equation}
 U_{I}\left(\tau_{1},\tau_{2}\right)=U_{0}^{S}(t_{0}^{+},\tau_{1})U^{S}\left(\tau_{1},\tau_{2}\right)U_{0}^{S}(\tau_{2},t_{0}^{+}),
\end{equation}
where $U_{0}^{S}$ is similar to $U^{S}$ but determined by $H_{0}^{x}\left(\tau\right)$.
According to the interaction-picture evolution operator, we can define the interaction-picture operator
such as $u_{C}^{I}\left(\tau_{1}\right)$ as
\begin{equation}
\label{interaction-picture operator_u_C}
u_{C}^{I}\left(\tau_{1}\right)=U_{0}^{S}\left(t_{0}^{+},\,\tau_{1}\right)u_{C}U_{0}^{S}\left(\tau_{1},\, t_{0}^{+}\right).
\end{equation}
The relation between the Heisenberg-picture operator and the interaction-picture one turns out to be
\begin{equation}
u_{C}^{H}\left(\tau_{1}\right)=U_{I}\left(t_{0}^{+},\tau_{1}\right)u_{C}^{I}\left(\tau_{1}\right)U_{I}\left(\tau_{1},t_{0}^{+}\right).
 \end{equation}

Further, the interaction-picture evolution operator can be expressed as
\begin{equation}
U_{I}\left(\tau_{1},\tau_{2}\right)=T_{\tau}e^{-\frac{i}{\hbar}\int_{C[\tau_{1},\tau_{2}]}H_{n}^{I}\left(\tau\right)d\tau}
\end{equation}
for $\tau_{1}$ succeeds $\tau_{2}$.
Using the interaction picture on the contour, we can rewrite the $\tilde{G}_{CC}$ from Eq~\eqref{Heisenberg_picture_G_CC} as
\begin{eqnarray}
   && \tilde{G}_{CC}\left(\tau_{1},\tau_{2}\right)=\nonumber \\
   &=& -\frac{i}{\hbar}\left\langle U_{0}^{S}\left(t_{0}^{-},t_{0}^{+}\right)T_{\tau}u_{C}^{I}\left(\tau_{1}\right)u_{C}^{I,T}\left(\tau_{2}\right)e^{-\frac{i}{\hbar}\int_{C}d\tau H_{n}^{I}\left(\tau\right)}\right\rangle \frac{1}{\mathcal{Z}},\nonumber\\
   \label{interaction picture_GCCC}
\end{eqnarray}
which is shown below assuming that  $\tau_{1}$ succeeds $\tau_{2}$ without loss of generality:
\begin{eqnarray}
\tilde{G}_{CC}\left(\tau_{1},\tau_{2}\right) & = & -\frac{i}{\hbar}\Big\langle U^{S}\left(t_{0}^{-},t_{M}^{-}\right)U^{S}\left(t_{M}^{+},t_{0}^{+}\right)\nonumber\\
&&u_{C}^{H}\left(\tau_{1}\right)u_{C}^{H,T}\left(\tau_{2}\right)\Big\rangle \frac{1}{\mathcal{Z}}\nonumber\\
 & = & -\frac{i}{\hbar}\bigg<U_{0}^{S}\left(t_{0}^{-},t_{0}^{+}\right)U_{I}\left(t_{0}^{-},t_{0}^{+}\right)U_{I}\left(t_{0}^{+},\tau_{1}\right)\nonumber\\
 &  &u_{C}^{I}\left(\tau_{1}\right)U_{I}\left(\tau_{1},t_{0}^{+}\right)U_{I}\left(t_{0}^{+},\tau_{2}\right)u_{C}^{I,T}\left(\tau_{2}\right)\nonumber\\
 &&U_{I}\left(\tau_{2},t_{0}^{+}\right)\bigg>\frac{1}{\mathcal{Z}}\nonumber\\
 & = & -\frac{i}{\hbar}\Big\langle U_{0}^{S}\left(t_{0}^{-},t_{0}^{+}\right)T_{\tau}u_{C}^{I}\left(\tau_{1}\right)u_{C}^{I,T}\left(\tau_{2}\right)\nonumber\\
 &&e^{-\frac{i}{\hbar}\int_{C}d\tau H_{n}^{I}\left(\tau\right)}\Big\rangle \frac{1}{\mathcal{Z}}.
\end{eqnarray}

By introducing
\[\mathcal{Z}_{0}=\Big\langle T_{\tau}e^{-\frac{i}{\hbar}\int_{C}d\tau\left(\hat{u}_{L}^{x,T}V^{LC}\hat{u}_{C}+\hat{u}_{C}^{T}V^{CR}\hat{u}_{R}\right)}\Big\rangle, \] which is the GF when $ H_n=0$, and defining $\mathcal{Z}_{n}=\mathcal{Z}/\mathcal{Z}_{0}$, $\tilde{G}_{CC}$ in Eq.~\eqref{interaction picture_GCCC} is written as
\begin{align}
 & \tilde{G}_{CC}\left(\tau_{1},\tau_{2}\right)=\nonumber \\
 & \!-\frac{i}{\hbar}\mathrm{Tr}\left[\rho_{ini}^{I}T_{\tau}u_{C}^{I}\left(\tau_{1}\right)u_{C}^{I,T}\left(\tau_{2}\right)e^{-\frac{i}{\hbar}\int_{C}d\tau H_{n}^{I}\left(\tau\right)}\right]\frac{1}{\mathcal{Z}_{n}}, \label{interaction picture_G_CC}
\end{align}
where $\rho_{ini}^{I}=\rho_{ini}U_{0}^{S}\left(t_{0}^{-},t_{0}^{+}\right)/\mathcal{Z}_{0}$, ($\mathrm{Tr}$$\left(\rho_{ini}^{I}\right)=1$).
Notice that $\rho_{ini}^{I}$ and the interaction-picture operator on the contour, such as $u_{C}^{I}\left(\tau_{1}\right)$, after second quantization satisfy the
sufficient conditions for the Wick theorem to be valid presented in the appendix~\ref{Wick}.

Observing the structure of Eq.~\eqref{interaction picture_G_CC} and realizing that the denominator $\mathcal{Z}_{n}$ cancels the disconnected diagrams,
we can obtain the Dyson equation for $\tilde{G}_{CC}$
as $\tilde{G}_{CC}=\tilde{G}_{CC}^{0}+\tilde{G}_{CC}^{0}\tilde{\Sigma}_{n}\tilde{G}_{CC}$,
a symbolic notation of
\begin{align}
\tilde{G}_{CC}\left(\tau_{1},\tau_{2}\right)= & \:\tilde{G}_{CC}^{0}\left(\tau_{1},\tau_{2}\right)\nonumber \\
+\qquad\quad\;\: &\!\!\!\!\!\!\!\!\!\!\!\!\!\!\!\!\!\!\!\!\int_{C}d\tau d\tau'\tilde{G}_{CC}^{0}\left(\tau_{1},\tau\right)\tilde{\Sigma}_{n}\left(\tau,\tau'\right)\tilde{G}_{CC}\left(\tau',\tau_{2}\right)\label{eq:GCC_tilt}
\end{align}
in terms of
\begin{align}
\tilde{G}_{CC}^{0}= & -\frac{i}{\hbar}\mathrm{Tr}\left[\rho_{ini}^{I}T_{\tau}u_{C}^{I}\left(\tau_{1}\right)u_{C}^{I,T}\left(\tau_{2}\right)\right]
\end{align}
and the nonlinear self energy $\tilde{\Sigma}_{n}$ constructed by the bare propagator
$\tilde{G}_{CC}^{0}$, whose vertices are solely due to the nonlinear Hamiltonian $H_{n}$.

Going to the interaction picture with respect to the free Hamiltonian $h=H_{L}+H_{C}+H_{R}$,
$\tilde{G}_{CC}^{0}$ can be written as
\begin{eqnarray}
   && \tilde{G}_{CC}^{0}\left(\tau_{1},\tau_{2}\right)=-\frac{i}{\hbar}\Big\langle T_{\tau}\hat{u}_{C}\left(\tau_{1}\right)\hat{u}_{C}^{T}\left(\tau_{2}\right)\nonumber\\
   && e^{-\frac{i}{\hbar}\int_{C}d\tau\hat{u}_{L}^{x,T}\left(\tau\right)V^{LC}\hat{u}_{C}\left(\tau\right)+\hat{u}_{C}^{T}\left(\tau\right)V^{CR}\hat{u}_{R}\left(\tau\right)}\Big\rangle \frac{1}{\mathcal{Z}_{0}}
\end{eqnarray}
so that
\begin{align}
\tilde{G}_{CC}^{0}= & \: g_{C}+g_{C}\left(\tilde{\Sigma}_{L}+\Sigma_{R}\right)\tilde{G}_{CC}^{0}\label{eq:GCC_0_tilt},
\end{align}
where $\Sigma_{R}\left(\tau_{1},\tau_{2}\right)$ is the right-lead
version of the ordinary contour-order self energy $\Sigma_{\nu}=V^{C\nu}g_{\nu}V^{\nu C},\,\nu=L,R$,
in which $g_{\alpha}\left(\tau_{1},\tau_{2}\right)_{jk}=-\frac{i}{\hbar}\left\langle T_{\tau}\hat{u}_{\alpha,j}\left(\tau_{1}\right)\hat{u}_{\alpha,k}\left(\tau_{2}\right)\right\rangle $
for $\alpha=L,C,R$ are the uncoupled contour-order Green's functions.

Though Eq.~\eqref{eq:GCC_tilt} and Eq.~\eqref{eq:GCC_tilt} are enough for the calculation of $\tilde{G}_{CC}$, for convenience
one can introduce an counting-field independent auxiliary equation $G_{CC}^{0}=g_{C}+g_{C}\left(\Sigma_{L}+\Sigma_{R}\right)G_{CC}^{0},$
and combine it with Eqs.~\eqref{eq:GCC_tilt} and \eqref{eq:GCC_0_tilt} to obtain
a closed Dyson equation for $\tilde{G}_{CC}\left(\tau_{1},\tau_{2}\right)$:
\begin{align}
\tilde{G}_{CC}= & \: G_{CC}^{0}+G_{CC}^{0}\left(\Sigma_{A}+\tilde{\Sigma}_{n}\right)\tilde{G}_{CC},\label{eq:Key_for_GCC_tilt}
\end{align}
where the shifted self energy $\Sigma_{A}\equiv\tilde{\Sigma}_{L}-\Sigma_{L}$,
which first appears in Ref.~\onlinecite{Wang2011}, accounts for the distribution of heat transfer
in ballistic systems.

From now on, for notational simplicity, all the subscripts $CC$ of
the Green's functions will be suppressed and the superscript $0$
in both $\tilde{G}_{CC}^{0}$ and $G_{CC}^{0}$ will be re-expressed
as a subscript.

Until now, the formalism for studying the distribution of heat transport
across general nonlinear junctions has been completely established.
In the case of steady state, one simply set $t_{0}\rightarrow-\infty$ and $t_{M}\rightarrow+\infty$
simultaneously, and technically assume that real-time versions of
$\tilde{G}\left(\tau_{1},\tau_{2}\right)$ are time-translationally
invariant. Then going to the Fourier space, Eq.~\eqref{generalized Meir_wingreen formula}
for $\frac{\partial\ln\mathcal{Z}}{\partial\left(i\xi\right)}$ in
steady state could be rewritten as
\begin{align}
\frac{\partial\ln\mathcal{Z}}{\partial\left(i\xi\right)}= & \left(t_{M}-t_{0}\right)\int_{-\infty}^{\infty}\!\! d\omega\frac{\hbar\omega}{2\pi}\mathrm{Tr}\left[\tilde{G}^{<}\Sigma_{L}^{>}e^{-i\hbar\omega\xi}\right]\label{eq:DLnZ_omega}
\end{align}
after taking into account $\tilde{G}^{>}\left[-\omega\right]=\tilde{G}^{<}\left[\omega\right]^{T}$
and $\Sigma_{L}^{<}\left[-\omega\right]=\Sigma_{L}^{>}\left[\omega\right]^{T}$.
In the Fourier space, due to Eq.~\eqref{eq:Key_for_GCC_tilt} exact
result for $\tilde{G}\left[\omega\right]$ could be yielded as
\begin{align}
\tilde{G}\left[\omega\right]= & \left(G_{0}\left[\omega\right]^{-1}-\Sigma_{A}\left[\omega\right]-\tilde{\Sigma}_{n}\left[\omega\right]\right)^{-1}\label{eq:G_tilt_omega}
\end{align}
when keeping in mind the convention that the contour-order Green's
function such as $\tilde{G}\left(\tau_{1},\tau_{2}\right)$ in frequency
space is written as
\begin{align}
\tilde{G}\left[\omega\right]= & \begin{bmatrix}\tilde{G}^{t}\left[\omega\right] & \tilde{G}^{<}\left[\omega\right]\\
-\tilde{G}^{>}\left[\omega\right] & -\tilde{G}^{\bar{t}}\left[\omega\right]
\end{bmatrix}.
\end{align}

\section{APPLICATION TO SINGLE-SITE JUNCTION\label{sec:APPLICATION}}
Now we apply the general formalism developed above to study a single-site junction with a quartic nonlinear on-site pinning potential, that
is, $H_{n}=\frac{1}{4}\lambda u_{C,0}^{4}$ in Eq.~\eqref{eq:total H}.
In this case, nonlinear contour-order self energy exact up to first
order in nonlinear strength is $\tilde{\Sigma}_{n}\left(\tau,\tau'\right)=3i\hbar\lambda\tilde{G}_{0}\left(\tau,\tau'\right)\delta\left(\tau,\tau'\right)$,
where the generalized $\delta$-function $\delta\left(\tau,\tau'\right)$
is the counterpart of the ordinary Dirac delta function on the contour
$C$, see, for example, Ref.~\onlinecite{Wang2008}. Thus the corresponding
frequency-space nonlinear self energy is
\begin{align}
\tilde{\Sigma}_{n}\left[\omega\right]= & \:3i\hbar\lambda\begin{bmatrix}\tilde{G}_{0}^{t}\left(0\right) & 0\\
0 & \tilde{G}_{0}^{\bar{t}}\left(0\right)
\end{bmatrix}.
\end{align}
Consequently, exact up to first order in nonlinear strength the CGF
for the molecular junction could be given as
\begin{align}
\frac{1}{\left(t_{M}-t_{0}\right)}\frac{\partial\ln\mathcal{Z}\left(\xi\right)}{\partial\left(i\xi\right)}= & -\int_{-\infty}^{\infty}\frac{d\omega}{4\pi}\Big\{\frac{\partial\ln D\left[\omega\right]}{\partial\left(i\xi\right)}-3i\hbar\lambda\nonumber \\
\times\Big[\tilde{G}_{0}^{t}\left(0\right)G_{0}^{t}\left[\omega\right]-\tilde{G}_{0}^{\bar{t}}\left(0\right) & \! G_{0}^{\bar{t}}\left[\omega\right]\Big]\frac{\partial}{\partial\left(i\xi\right)}\frac{1}{D\left[\omega\right]}\Big\}\label{eq:LnZ_molecular}
\end{align}
with
\begin{align}
& D[\omega]\equiv \det\Big[I-G_{0}\left[\omega\right]\Sigma_{A}\left[\omega\right]\Big]\nonumber \\
&=1\!-\! T[\omega]\!\Big[\!\left(e^{i\xi\hbar\omega}\!-\!1\right)\!f_{L}\!\left(1\!+\!f_{R}\right)\!+\!\left(e^{\!-i\xi\hbar\omega}\!-\!1\right)\!f_{R}\!\left(1\!+\!f_{L}\right)\!\Big]
\end{align}
and $\tilde{G}_{0}^{t,\bar{t}}\left(0\right)=\int_{-\infty}^{\infty}\frac{d\omega}{2\pi}G_{0}^{t,\bar{t}}\left[\omega\right]/D\left[\omega\right]$,
where $T\left[\omega\right]=\mathrm{Tr}\left(G_{0}^{r}\Gamma_{R}G_{0}^{a}\Gamma_{L}\right)$
is the transmission coefficient in the ballistic system, and $f_{\left\{ L,R\right\} }=\left\{ \exp\left(\beta_{\{L,R\}}\hbar\omega\right)-1\right\} ^{-1}$
is the Bose-Einstein distribution function for phonons. Here $G_{0}^{r}=G_{0}^{t}-G_{0}^{<}$
and $G_{0}^{a}=G_{0}^{<}-G_{0}^{\bar{t}}$ are retarded and advanced
Green's functions, respectively. Also $\Gamma_{\left\{ L,R\right\} }=i\left[\Sigma_{\left\{ L,R\right\} }^{r}-\Sigma_{\left\{ L,R\right\} }^{a}\right]$,
related to the spectral density of the baths, are expressed by retarded
and advanced self energies similarly defined as Green's functions. Eq.~\eqref{eq:LnZ_molecular} satisfies Gallavotti-Cohen symmetry
\cite{Gallavotti1995} for the derivatives, since $D\left[\omega\right]$
remains invariant under the transformation $\xi\rightarrow-\xi+i\left(\beta_{R}-\beta_{L}\right)$
while $\partial D\left[\omega\right]/\partial\left(i\xi\right)$ changes
sign.

One could easily use this CGF in Eq.~\eqref{eq:LnZ_molecular} to
evaluate cumulants. The steady current out of the left lead is closely
related to the first cumulant so that
\begin{align}
I^{ss}= & \frac{d}{dt_{M}}\left(\frac{\partial\mathcal{\ln Z}\left(\xi\right)}{\partial\left(i\xi\right)}\bigg|_{\xi=0}\right)\nonumber \\
= & \int_{-\infty}^{\infty}\frac{d\omega}{4\pi}\hbar\omega\left(1+\Lambda\left[\omega\right]\right)T\left[\omega\right]\left(f_{L}-f_{R}\right),
\end{align}
where $\Lambda\left[\omega\right]\equiv3i\hbar\lambda G_{0}^{t}\left(0\right)\left(G_{0}^{a}\left[\omega\right]+G_{0}^{r}\left[\omega\right]\right)$
is the first-order nonlinear correction to the transmission coefficient.

The fluctuation for steady-state heat transfer in the molecular junction
is obtained by taking the second derivative with respect to $i\xi$,
and then setting $\xi=0$:
\begin{eqnarray}
\frac{\left\langle \left\langle Q^{2}\right\rangle \right\rangle}{\left(t_{M}-t_{0}\right)}&=&\!\int_{-\infty}^{\infty}\!\frac{d\omega}{4\pi}\Big\{(\hbar\omega)^{2} T^{2}[\omega] (1+2\Lambda[\omega]) \left(f_{L}-f_{R}\right)^{2} \nonumber \\
&+& 3\hbar^{2}\lambda\omega\left[G_{0}^{\bar{t}}[\omega] \delta \tilde{G}_{0}^{\bar{t}} \!-\!G_{0}^{t}[\omega] \delta \tilde{G}_{0}^{t} \right] T[\omega] \left(f_{L}-f_{R}\right) \nonumber \\
&+&\left(\hbar\omega\right)^{2} T[\omega](1+\Lambda[\omega])(f_L\!+\!f_R\!+\!2f_L f_{R})\Big\},
\end{eqnarray}
where,
\begin{equation}
\delta \tilde{G}_{0}^{t,{\bar{t}}} \equiv \frac{\partial\tilde{G}_{0}^{\bar{t},t}\left(0\right)}{\partial\xi}\bigg|_{\xi=0}\!=\!-i\int_{-\infty}^{\infty}\frac{d\omega}{2\pi}\, \hbar\omega\, T[\omega]\left(f_{L}-f_{R}\right)G_{0}^{\bar{t},t}[\omega].
\end{equation}

Higher-order cumulants can be also systematically given by corresponding
higher-order derivatives, although the expressions are messy.

\begin{figure}
\includegraphics[width=\columnwidth]{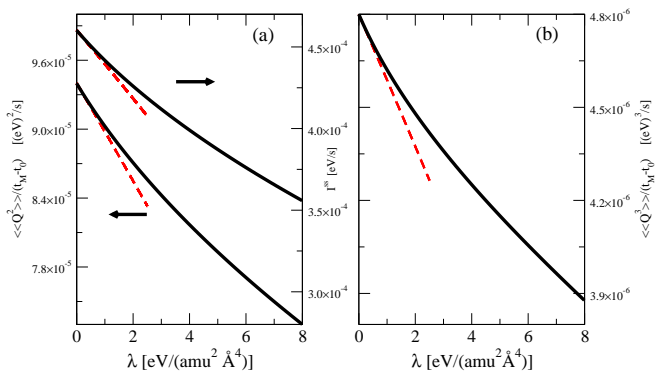}
\caption[The first three steady-state cumulants with nonlinear
strength]{The first three steady-state cumulants with nonlinear
strength $\lambda$ for $k=1\,\mathrm{eV/\left(u\AA^{2}\right)}$,
$k_{0}=0.1k$, $K_{C}=1.1k$, and $V_{-1,0}^{LC}=V_{0,1}^{CR}=-0.25k$.
The solid (dotted) line shows the self-consistent (first-order in
$\lambda$) results for the cumulants. The temperatures of the left
and right lead are 660 K and 410 K, respectively.\label{fig 1}}

\end{figure}

In Fig~\ref{fig 1}, we give a numerical illustration to the first
three cumulants for heat transfer in this molecular junction using a
self-consistent procedure~\cite{Park2011}, which means that the nonlinear
contour-order self energy is taken as $\tilde{\Sigma}_{n}\left(\tau,\tau'\right)=3i\hbar\lambda\tilde{G}\left(\tau,\tau'\right)\delta\left(\tau,\tau'\right)$.
Very recently, it is shown that such self-consistent calculation gives
extremely accurate results for the current in the case of a single
site model as compared with master equation approach,~\cite{Thingna2012}
thus we believe that it should leads to excellent predictions for
the FCS.

As shown, the effect of nonlinearity is
to reduce the current as well as higher order fluctuations, and the
fact that third and higher order cumulants are small but nonzero implies
that the distribution for transferred energy is not Gaussian. In this
numerical illustration, the Rubin baths are used, that is, $K_{\alpha},\,\alpha=L,\, R$
in Eq.~\eqref{eq:total H} are both the semi-infinite tridiagonal
spring constant matrix consisting of $2k+k_{0}$ along the diagonal
and $-k$ along the two off-diagonals. And only the nearest interaction
$V_{-1,0}^{LC}$ and $V_{0,1}^{CR}$ between the molecular and the
two bathes are considered and $H_{C}=\frac{1}{2}p_{C,0}^{2}+\frac{1}{2}K_{C}u_{C,0}^{2}$.
As expected, for weak nonlinearity the first-order perturbation results, presented
as dotted lines, are consistent with the corresponding self-consistent
ones.

\section{Summary\label{sec:SUMMARY}}

A formally rigorous formalism dealing with cumulants of heat transfer
across nonlinear quantum junctions is established based on field
theoretical and NEGF methods. The CGF for the heat transfer in both
transient and steady-state regimes is studied on an equal footing
and useful formulas for the CGF are obtained. A new feature of this
formalism is that counting-field dependent full Green's function $\tilde{G}_{CC}$ can
be expressed solely through the nonlinear term $H_{n}^{I}\left(\tau\right)$
with the help of an interaction-picture transformation defined on a contour. Although
we focus on the distribution of heat transfer in pure nonlinear phononic systems, there is no doubt that this
general formalism can be readily employed to handle any other nonlinear
systems, such as electron-phonon interaction
and Joule heating problems. Up to the first order in the nonlinear strength for the single-site
quartic model, the CGF for steady-state heat transfer is obtained
and explicit results for the steady current and fluctuation of steady-state
heat transfer are given. A self-consistent procedure, which works
well for strong nonlinearity, is also introduced to numerically check our general
formalism.

\begin{acknowledgments}
We thank Juzar Thingna, Lifa Zhang and Sha Liu for insightful discussions.
This work is supported in part by URC Grant No. R-144-000-285-646, and R-144-000-300-112.
\end{acknowledgments}

\appendix
\section{The Wick Theorem (Phonons)}
\label{Wick}

In this appendix, we give sufficient conditions for the Wick
theorem to be valid which covers most of the situation we encounter.
The discussion is limited to the case of bosonic operators which is
the main interest in this paper. We mainly extend Gaudin's approach.~\cite{Gaudin1960}
For an alternative proof, one can resort ro the Ref.~\onlinecite{Blaizot1986}.

First we explain what the Wick theorem is. The Wick theorem says that
the average value of a product of creation and annihilation operators
is equal to the sum of all complete systems of pairings, mathematically
which can be stated as
\begin{eqnarray}
&&\mathrm{Tr}\left\{ \rho^{ini}\beta_{1}\beta_{2}\cdots\beta_{s}\right\}\nonumber\\
  & = & \mathrm{Tr}\left\{ \rho^{ini}\beta_{1}\beta_{2}\right\} \mathrm{Tr}\left\{ \rho^{ini}\beta_{3}\beta_{4}\cdots\beta_{s}\right\} \nonumber \\
 & + & \mathrm{Tr}\left\{ \rho^{ini}\beta_{1}\beta_{3}\right\} \mathrm{Tr}\left\{ \rho^{ini}\beta_{2}\beta_{4}\cdots\beta_{s}\right\} \label{eq:Wick's Theorem}\\
 & + & \cdots\nonumber \\
 & + & \mathrm{Tr}\left\{ \rho^{ini}\beta_{1}\beta_{s}\right\} \mathrm{Tr}\left\{ \rho^{ini}\beta_{2}\beta_{3}\cdots\beta_{s-1}\right\} \nonumber
\end{eqnarray}
and then applying this relation recursively to all of the multiple
operator averages until only pairs of operators remain.

Now we explore the sufficient conditions for the Wick theorem to be
justified, which simply means that Eq. (\ref{eq:Wick's Theorem})
is valid. Suppose the system's degrees of freendom is $f$, and we define
\begin{equation}
\label{alpha}
\alpha=\begin{pmatrix}a\\
a^{\dagger}
\end{pmatrix},\:\alpha_{i}=a_{i},\:\alpha_{f+i}=a_{i}^{\dagger},\: i=1,2,\ldots f,
\end{equation}
where $a_{i}$ and $a_{i}^{\dagger}$ are annihilation and creation
operators respectively.

Assume
\begin{eqnarray}
\alpha_{i}\rho^{ini} & = & \sum_{k=1}^{2f}h_{ik}\rho^{ini}\alpha_{k},\label{eq:condition1}
\end{eqnarray}
where $h_{ik}$ are $c$-numbers.
We prove the Wick theorem for $\mathrm{Tr}\left\{ \rho^{ini}\alpha_{i_{1}}\alpha_{i_{2}}\cdots\alpha_{i_{s}}\right\} $,
which is shown below:
\begin{eqnarray}
&&\mathrm{Tr}\left\{ \rho^{ini}\alpha_{i_{1}}\alpha_{i_{2}}\cdots\alpha_{i_{s}}\right\} \nonumber\\
 & = & \mathrm{Tr}\left\{ \rho^{ini}\left[\alpha_{i_{1}},\:\alpha_{i_{2}}\right]\cdots\alpha_{i_{s}}\right\} +\mathrm{Tr}\left\{ \rho^{ini}\alpha_{i_{2}}\alpha_{i_{1}}\cdots\alpha_{i_{s}}\right\} \nonumber\\
 & = & \mathrm{Tr}\left\{ \rho^{ini}\left[\alpha_{i_{1}},\:\alpha_{i_{2}}\right]\cdots\alpha_{i_{s}}\right\} +\mathrm{Tr}\left\{ \rho^{ini}\alpha_{i_{2}}\left[\alpha_{i_{1}},\:\alpha_{i_{3}}\right]\cdots\alpha_{i_{s}}\right\} \nonumber\\
 & +& \mathrm{Tr}\left\{ \rho^{ini}\alpha_{i_{2}}\alpha_{i_{3}}\alpha_{i_{1}}\cdots\alpha_{i_{s}}\right\} \nonumber\\
 & = & \mathrm{Tr}\left\{ \rho^{ini}\left[\alpha_{i_{1}},\:\alpha_{i_{2}}\right]\cdots\alpha_{i_{s}}\right\} +\mathrm{Tr}\left\{ \rho^{ini}\alpha_{i_{2}}\left[\alpha_{i_{1}},\:\alpha_{i_{3}}\right]\cdots\alpha_{i_{s}}\right\} \nonumber\\
 &  +& \mathrm{Tr}\left\{ \rho^{ini}\alpha_{i_{2}}\alpha_{i_{3}}\left[\alpha_{i_{1}},\:\alpha_{i_{4}}\right]\cdots\alpha_{i_{s}}\right\}\nonumber\\
 &+&\mathrm{Tr}\left\{ \rho^{ini}\alpha_{i_{2}}\alpha_{i_{3}}\alpha_{i_{4}}\alpha_{i_{1}}\cdots\alpha_{i_{s}}\right\} \nonumber\\
 & = & \cdots\nonumber\\
 & = & \sum_{j=2}^{s}\left[\alpha_{i_{1}},\:\alpha_{i_{j}}\right]\mathrm{Tr}\left\{ \rho^{ini}\overset{\circ}{\alpha}_{i_{1}}\alpha_{i_{2}}\cdots\overset{\circ}{\alpha}_{i_{j}}\cdots\alpha_{i_{s}}\right\}\nonumber \\
 &  +& \mathrm{Tr}\left\{ \alpha_{i_{1}}\rho^{ini}\alpha_{i_{2}}\alpha_{i_{3}}\cdots\alpha_{i_{s}}\right\}\nonumber \\
 & = & \sum_{j=2}^{s}\left[\alpha_{i_{1}},\:\alpha_{i_{j}}\right]\mathrm{Tr}\left\{ \rho^{ini}\overset{\circ}{\alpha}_{i_{1}}\alpha_{i_{2}}\cdots\overset{\circ}{\alpha}_{i_{j}}\cdots\alpha_{i_{s}}\right\} \nonumber\\
 &  +& \sum_{k=1}^{2f}h_{i_{1}k}\mathrm{Tr}\left\{ \rho^{ini}\alpha_{k}\alpha_{i_{2}}\alpha_{i_{3}}\cdots\alpha_{i_{s}}\right\}
\end{eqnarray}
where the circle over the operator means that this operator is omitted.
Then
\begin{eqnarray*}
   &=& \sum_{k=1}^{2f}\left(1-h\right)_{i_{1}k}\mathrm{Tr}\left\{ \rho^{ini}\alpha_{k}\alpha_{i_{2}}\cdots\alpha_{i_{s}}\right\}  \\
   &=& \sum_{j=2}^{s}\left[\alpha_{i_{1}},\:\alpha_{i_{j}}\right]\mathrm{Tr}\left\{ \rho^{ini}\overset{\circ}{\alpha}_{i_{1}}\alpha_{i_{2}}\cdots\overset{\circ}{\alpha}_{i_{j}}\cdots\alpha_{i_{s}}\right\}
\end{eqnarray*}
Multiply by the inverse matrix $\left(1-h\right)^{-1}$, we can get
\begin{align}
 & \mathrm{Tr}\left\{ \rho^{ini}\alpha_{i_{1}}\alpha_{i_{2}}\cdots\alpha_{i_{s}}\right\} \nonumber \\
=\;\; &\!\!\! \!\!\!\sum_{j=2}^{s}\left\{ \sum_{k=1}^{2f}\left(1-h\right)_{i_{1}k}^{-1}\left[\alpha_{k},\:\alpha_{i_{j}}\right]\right\}\!\! \mathrm{Tr}\left\{ \rho^{ini}\overset{\circ}{\alpha}_{i_{1}}\alpha_{i_{2}}\cdots\overset{\circ}{\alpha}_{i_{j}}\cdots\alpha_{i_{s}}\right\} \label{eq:intermidiate}
\end{align}
After considering the special case
\begin{equation}
\mathrm{Tr}\left\{ \rho^{ini}\alpha_{i_{1}}\alpha_{i_{j}}\right\} =\sum_{k=1}^{2f}\left(1-h\right)_{i_{1}k}^{-1}\left[\alpha_{k},\:\alpha_{i_{j}}\right],
\end{equation}
we obtain from Eq.~(\ref{eq:intermidiate})
\begin{eqnarray*}
   && \mathrm{Tr}\left\{ \rho^{ini}\alpha_{i_{1}}\alpha_{i_{2}}\cdots\alpha_{i_{s}}\right\} \\
   &=& \sum_{j=2}^{s}\mathrm{Tr}\left\{ \rho^{ini}\alpha_{i_{1}}\alpha_{i_{j}}\right\} \mathrm{Tr}\left\{ \rho^{ini}\overset{\circ}{\alpha}_{i_{1}}\alpha_{i_{2}}\cdots\overset{\circ}{\alpha}_{i_{j}}\cdots\alpha_{i_{s}}\right\},
\end{eqnarray*}

Assume
\begin{equation}
\beta_{j}=\sum_{i=1}^{2f}g_{ji}\alpha_{i}\label{eq:condition2}
\end{equation}
where $g_{ji}$ are $c$-numbers. Then
\begin{eqnarray*}
&&\mathrm{Tr}\left\{ \rho^{ini}\beta_{1}\beta_{2}\cdots\beta_{s}\right\} \\
 & = & \sum_{i_{1}}\sum_{i_{2}}\cdots\sum_{i_{s}}g_{1i_{1}}g_{2i_{2}}\cdots g_{si_{2}}\mathrm{Tr}\left\{ \rho^{ini}\alpha_{i_{1}}\alpha_{i_{2}}\cdots\alpha_{i_{s}}\right\} \\
 & = & \sum_{i_{1}}\sum_{i_{2}}\cdots\sum_{i_{s}}g_{1i_{1}}g_{2i_{2}}\cdots g_{si_{2}}\sum_{j=2}^{s}\mathrm{Tr}\left\{ \rho^{ini}\alpha_{i_{1}}\alpha_{i_{j}}\right\} \\
 &  & \times\mathrm{Tr}\left\{ \rho^{ini}\overset{\circ}{\alpha}_{i_{1}}\alpha_{i_{2}}\cdots\overset{\circ}{\alpha}_{i_{j}}\cdots\alpha_{i_{s}}\right\} \\
 & = & \sum_{j=2}^{s}\mathrm{Tr}\left\{ \rho^{ini}\beta_{1}\beta_{j}\right\} \mathrm{Tr}\left\{ \rho^{ini}\overset{\circ}{\beta}_{1}\beta_{2}\cdots\overset{\circ}{\beta}_{j}\cdots\beta_{s}\right\}
\end{eqnarray*}
which is just the Eq.~(\ref{eq:Wick's Theorem}).

In summary, the sufficient conditions for the Wick theorem Eq.~(\ref{eq:Wick's Theorem})
to be valid are Eq.~(\ref{eq:condition1}) and Eq.~(\ref{eq:condition2})
and implicitly $\mathrm{Tr}\left(\rho^{ini}\right)=1$.

In the following, we try to figure out the form of initial density
matrix $\rho^{ini}$ satisfying Eq.~(\ref{eq:condition1}), which
turns out to be
\begin{equation}
\rho^{ini}=e^{-\alpha^{T}A\alpha}\label{eq:quadratic}
\end{equation}
with $A$ to be a general square matrix. We neglect the normalization
constant for $\mathrm{Tr}\left(\rho^{ini}\right)=1$ here.
To this end, we split the $A$ to be a symmetrical part and an anti-symmetrical
part, that is
\begin{eqnarray}
A & = & \frac{1}{2}\left(A+A^{T}\right)+\frac{1}{2}\left(A-A^{T}\right)\\
 & \equiv & A^{s}+A^{a}.
\end{eqnarray}
Let us define
\begin{eqnarray*}
f_{i}\left(t\right) & \equiv & e^{t\alpha^{T}A\alpha}\alpha_{i}e^{-t\alpha^{T}A\alpha}\\
 & = & e^{t\frac{1}{2}\alpha^{T}A^{s}\alpha}\alpha_{i}e^{-t\frac{1}{2}\alpha^{T}A^{s}\alpha}.
\end{eqnarray*}
In obtaining the second equality, notice that $\alpha^{T}A^{a}\alpha$
is a $c$-number due to $\left[\alpha,\,\alpha^{\dagger}\right]=\begin{pmatrix}1 & 0\\
0 & -1
\end{pmatrix}$ and $A^{a,T}=-A^{a}$.
Thus
\begin{eqnarray*}
\frac{df_{i}\left(t\right)}{dt} & = & e^{t\frac{1}{2}\alpha^{T}A^{s}\alpha}\left[\frac{1}{2}\alpha^{T}A^{s}\alpha,\,\alpha_{i}\right]e^{-t\frac{1}{2}\alpha^{T}A^{s}\alpha}\\
 & = & -\sum_{j}\left(\sigma A^{s}\right)_{ij}f_{j}\left(t\right),
\end{eqnarray*}
where $\sigma\equiv\begin{pmatrix}0 & -1\\
1 & 0
\end{pmatrix}$.
So $f_{i}\left(t\right)=\sum_{j}\left(e^{-t\sigma A^{s}}\right)_{ij}\alpha_{j}$
and $f_{i}\left(1\right)=e^{\alpha^{T}A\alpha}\alpha_{i}e^{-\alpha^{T}A\alpha}=\sum_{j}\left(e^{-\sigma A^{s}}\right)_{ij}\alpha_{j}$
or equivalently
\begin{equation}
\alpha_{i}e^{-\alpha^{T}A\alpha}=\sum_{j}\left(e^{-\sigma A^{s}}\right)_{ij}e^{-\alpha^{T}A\alpha}\alpha_{j}.
\end{equation}
More generally, the multiplication of finite number of the form of
Eq.~(\ref{eq:quadratic}) still satisfies Eq.~(\ref{eq:condition1}),
such as
\begin{equation}
\rho^{ini}=e^{-\alpha^{T}A\alpha}e^{-\alpha^{T}B\alpha},
\end{equation}
which is shown below:
\begin{eqnarray*}
\alpha_{i}\rho^{ini} & = & \alpha_{i}e^{-\alpha^{T}A\alpha}e^{-\alpha^{T}B\alpha}\\
 & = & \sum_{j}\left(e^{-\sigma A^{s}}\right)_{ij}e^{-\alpha^{T}A\alpha}\alpha_{j}e^{-\alpha^{T}B\alpha}\\
 & = & \sum_{j}\left(e^{-\sigma A^{s}}\right)_{ij}e^{-\alpha^{T}A\alpha}\sum_{k}\left(e^{-\sigma B^{s}}\right)_{jk}e^{-\alpha^{T}B\alpha}\alpha_{k}\\
 & = & \sum_{j}\sum_{k}\left(e^{-\sigma A^{s}}\right)_{ij}\left(e^{-\sigma B^{s}}\right)_{jk}\rho^{ini}\alpha_{k}\\
 & = & \sum_{k}\left(e^{-\sigma A^{s}}e^{-\sigma B^{s}}\right)_{ik}\rho^{ini}\alpha_{k}.
\end{eqnarray*}

Due to the sufficient conditions presented in this appendix, the Wick theorem used in this paper for the Feynman-diagrammatic analysis is justified.
For example, for the case of the interaction picture on the contour, initial density matrix  $\rho_{ini}^{I}=\rho_{ini}U_{0}^{S}\left(t_{0}^{-},t_{0}^{+}\right)/\mathcal{Z}_{0}$$=\Pi_{\alpha=L,C,R}\frac{e^{-\beta_{\alpha}H_{\alpha}}}{\mathrm{Tr}\left(e^{-\beta_{\alpha}H_{\alpha}}\right)}e^{-\frac{i}{\hbar}H_0^x(t^-)(t_0-t_M)}e^{-\frac{i}{\hbar}H_0^x(t^+)(t_M-t_0)}/\mathcal{Z}_{0}$
is the multiplication of finite number of the form of
Eq.~(\ref{eq:quadratic}) and $\mathrm{Tr}$$\left(\rho_{ini}^{I}\right)=1$.
In addition, interaction-picture operator on the contour such as $u_{C}^{I}\left(\tau_{1}\right)$ in the Eq.~\eqref{interaction-picture operator_u_C} can be expressed as the linear transformation of $\alpha$ defined in the Eq.~\eqref{alpha} according to the similar steps for the calculation of $f_i(t)$.


\begin{thebibliography}{99}%
\bibitem{Blanter2000} Y. M. Blanter and M. B\"uttiker, Phys. Rep. \textbf{336}, 1, (2000).
\bibitem{Levitov1993} L. S. Levitov and G. B. Lesovik, JETP Lett. \textbf{58}, 230 (1993).
\bibitem{Saito2007} K. Saito and A. Dhar, Phys. Rev. Lett. \textbf{99}, 180601 (2007).
\bibitem{Schonhammer2007} K. Sch\"onhammer, Phys. Rev. B \textbf{75}, 205329 (2007).
\bibitem{Wang2011}
J.-S. Wang, B. K. Agarwalla, and H. Li, Phys. Rev. B \textbf{84}, 153412, (2011); B. K. Agarwalla, B. Li, and J.-S. Wang, Phys. Rev. E \textbf{85}, 051142 (2012).

\bibitem{Avriller2009} R. Avriller and A. Levy Yeyati, Phys. Rev. B \textbf{80}, 041309 (2009).
\bibitem{Wu2009} L.-A. Wu and D. Segal, Phys. Rev. Lett. \textbf{102}, 095503 (2009); C. W. Chang, D. Okawa, A. Majumdar, and A. Zettl, Science \textbf{314}, 1121 (2006).
\bibitem{Li2012} N. Li, J. Ren, L. Wang, G. Zhang, P. H\"anggi, and B. Li, Rev. Mod. Phys. \textbf{84}, 1045 (2012).
\bibitem{Ren2012}  J. Ren, P. H\"anggi, and B. Li, Phys. Rev. Lett. \textbf{104}, 170601 (2010); J. Ren, S. Liu, and B. Li, Phys. Rev. Lett. \textbf{108}, 210603 (2012).
\bibitem{Esposito2009} M. Esposito, U. Harbola, and S. Mukamel, Rev. Mod. Phys. \textbf{81}, 1665 (2009).

\bibitem{Griffiths2002} R. B. Griffiths, \textit{Consistent Quantum Theory}, Cambridge Univ. Press, Cambridge (2002).
\bibitem{Li_CGF} H. Li, B. K. Agarwalla, and J.-S. Wang, Phys. Rev. B \textbf{86}, 165425 (2012).
\bibitem{Gogolin2006} A. O. Gogolin and A. Komnik, Phys. Rev. B \textbf{73}, 195301 (2006).
\bibitem{Feynman1939} R. P. Feynman, Phys. Rev. \textbf{56}, 340 (1939).
\bibitem{Rammer2007} J. Rammer, \textit{Quantum Field Theory of Nonequilibrium States} (Cambridge 2007).
\bibitem{Kleinert2000} H. Kleinert, A. Pelster, B. Kastening, and M. Bachmann, Phys. Rev. E \textbf{62}, 1537 (2000).
\bibitem{Pelster2004} A. Pelster and K. Glaum, Phys. A \textbf{335}, 455 (2004).
\bibitem{Wang2008} J.-S. Wang, J. Wang, and J. T. L\"u,
Eur. Phys. J. B \textbf{62}, 381 (2008); J.-S. Wang, B. K. Agarwalla, H. Li, and J. Thingna, arxiv: 1303.7317.

\bibitem{Gallavotti1995} G. Gallavotti and E. G. D. Cohen. Phys. Rev. Lett. \textbf{74}, 2694 (1995).
\bibitem{Park2011} T.-H. Park and M. Galperin, Phys. Rev. B \textbf{84}, 205450 (2011).
\bibitem{Thingna2012} D. He and J. Thingna, private communication, see also J. Thingna, J. L. Garc\'{i}a-palacios, and J.-S. Wang, Phys. Rev. B \textbf{85}, 195452 (2012).
    \bibitem{Gaudin1960} M. Gaudin, Nucl. Phys.\textbf{15}, 89 (1960).
\bibitem{Blaizot1986} J.-P. Blaizot and G. Ripka, \textit{Quantum Theory of Finite Systems } (Massachusetts: MIT Press, 1986).
\end{thebibliography}
\end{document}